\documentclass{article}%
\usepackage{amsmath}
\usepackage{amsfonts}
\usepackage{amssymb}
\usepackage{graphicx}%
\newcommand{\bpartial}{\mathop{\partial\kern -4pt\raisebox{.8pt}{$|$}}}
\newcommand{\bra}{\mathopen{[\kern-1.6pt[}}
\newcommand{\ket}{\mathclose{]\kern-1.5pt]}}
\newcommand{\bbra}{\mathopen{[\kern-2.2pt[\kern-2.3pt[}}
\newcommand{\bket}{\mathclose{]\kern-2.1pt]\kern-2.3pt]}}

\makeindex
\usepackage{color}

\begin{document}

\title {\large{ \bf Gravity and induced matter on Nearly K\"{a}hler Manifolds }}

\vspace{3mm}

\author {  \small{ \bf  F. Naderi}\hspace{-1mm}{ \footnote{ e-mail: f.naderi@azaruniv.edu}} ,{ \small
} \small{ \bf  A. Rezaei-Aghdam}\hspace{-1mm}{
\footnote{Corresponding author. e-mail:
rezaei-a@azaruniv.edu}} \,and \small{ \bf  F. Darabi}\hspace{-1mm}{
\footnote{e-mail:
f.darabi@azaruniv.edu}}\\
{\small{\em Department of Physics, Faculty of Science, Azarbaijan Shahid Madani University}}\\
{\small{\em   53714-161, Tabriz, Iran  }}}

\maketitle

\begin{abstract}
We show that the conservation of energy-momentum tensor of a gravitational model with Einstein-Hilbert like action on a nearly K\"{a}hler manifold with the scalar curvature of a curvature-like tensor, is consistent with the nearly K\"{a}hler properties. In this way, the nearly K\"{a}hler structure is automatically manifested in the action as a induced matter field. As an example of nearly K\"{a}hler manifold, we consider the group manifold of $R\times II \times S^{3}\times S^{3}$ on which we identify a nearly K\"{a}hler structure and solve the Einstein equations to interpret the model. It is shown that the nearly K\"{a}hler structure in this example is capable of alleviating the well
known fine tuning  problem of the cosmological constant. Moreover, this structure may be considered as a potential candidate for dark energy.  
\end{abstract}
\section{Introduction}
Almost complex structure has been employed in the study of superstring theory, gravity and sigma model in different but related researches. In topological sigma model developed by Witten \cite{witten}, almost complex structure arose in a coupling term with sigma model. Following this and the idea of complexifying space-time, an invariant action containing exterior derivative of almost complex structure was constructed by Chamseddine \cite{ch} which gives the correct equation of motion of a complex metric in linearized limit\footnote{The complex metric is based on Einstein's attempts to generalize the relativistic theory of gravitation to establish a unified
field theory \cite {eins1, eins2}.}.  Other applications of the almost complex structure are in string theory (see for example \cite{Strominger}), where a set of 10-dimensional solution of string equation is based on complex non-K\"{a}hler manifolds.

 An interesting class of  non-K\"{a}hler manifolds is nearly K\"{a}hler manifolds. These manifolds were first studied  by A. Gray \cite{gary1,gary2,gary3}, and recently were investigated and classified by Nagy  {\it et al} \cite{nagy1, nagy4} who have shown that the complete and strict nearly K\"{a}hler manifolds, i.e. non-K\"{a}hler manifolds, are locally Riemannian products of 6-dimensional nearly K\"{a}hler manifolds, twistor spaces over quaternionic K\"{a}hler manifolds and homogeneous nearly K\"{a}hler spaces. The only known compact strict nearly K\"{a}hler manifolds in dimension 6 are three coset space 
$S^{6}\simeq G(3)/SU(3),  CP^{3}\simeq Sp(2)/SU(2)\times U(1),  F(1,2)\simeq SU(3)/ U(1)\times U(1)$ and a group manifold $S^{3}\times S^{3}\simeq SU(2)\times SU(2)$ \cite{Butruille}. Nearly K\"{a}hler manifolds have been of recent interest in massive type IIA super-gravity and related Yang-Mills theory, M-theory and heterotic strings compactification \cite{app}.

Besides, as an interesting appearance of almost complex structure in general relativity, in Ref.\cite{cfkaluzaklein} it has been shown that a special class of solutions of Kaluza-Klein conformal flat reduction equations relates the Kaluza-Klein gauge field $F_{\mu\nu}$ to the pseudo-K\"{a}hler and para-K\"{a}hler structure on manifold.

Here, we are interested in a somewhat different approach to develop a gravitational model which depends on almost complex 
structure. Following Einstein realization which declare that gravity should be regarded as a property of Riemannian geometry and 
space-time, it is intriguing to ask if another geometrical structure on manifold could play a physical role, for example, 
as a matter field. The main purpose of this paper is to construct an action of type $S(g_{MN},J_{M}^{~N}) $ on a non-K\"{a}hler 
manifold which includes not higher than second derivative of metric. The model uses a curvature-like tensor including the almost complex structure beside the metric structure  and the idea behind this is to give matter interpretation to the almost complex structure. Explicitly, the four dimensional matter will be induced from the
almost complex structure in accordance with the Einstein's dream that the
origin of matter is geometry.
 {In general, the manifold here is considered  to be non-K\"{a}hlerian,  however, it turns out that if one is interested in exploiting   an interpretation of matter source from the almost complex structure, the only choice is the nearly K\"{a}hler manifold which is  consistent with the conservation law for such a matter source}. In other words, conservation of energy-momentum tensor of model requires the manifold to be nearly 
K\"{a}hler. This type of manifolds appears 
in string compactification as internal space as a result of supersymmetry condition \cite{app}. Of course, in some previous works the authors have tried to include the complex 
structure by adding terms to the standard Einstein-Hilbert action (see for example Ref. \cite {ch}). In the present model, we will show that  such 
terms in the action are recovered in a straightforward way by using a Einstein-Hilbert action whose scalar curvature is constructed by the 
curvature-like tensor. We will show that such a new geometric structure is also capable of alleviating the well known fine tuning problem of the cosmological 
constant,  in
a typical example. Moreover, it may shed light on the other problems of cosmology
like dark energy.

The plan of this paper is as follows. A short summary of nearly K\"{a}hler manifolds is presented in section two. In section three, by investigation of symmetry properties of a tensor which carries almost complex structure  it is shown that such tensor is a curvature-like tensor. In section four, a gravitational model is constructed by the scalar curvature of this curvature-like tensor subject to the weak nearly  K\"{a}hlerian property imposed as a condition by a Lagrange multiplier. Then it is shown that the conservation of  energy-momentum tensor is consistent with the strong nearly K\"{a}hlerian condition. In section five, the Einstein field equations with the nearly K\"{a}hlerian properties of almost complex structure are solved, for example, on the group manifold $R\times II \times S^{3}\times S^{3}$,  the gravitational system (metric and matter) is completely determined, and a solution is given for the fine
tuning problem of the cosmological constant. In section six, we discuss on the possible role of almost complex structure as  dark energy. The paper ends with a conclusion.
\section{Nearly K\"{a}hler manifolds}
Here, for self containing of the paper we review the definitions and some mathematical concepts of nearly K\"{a}hler manifolds. Let $M$ be an almost Hermitian manifold with real dimension $d$ $(d>2)$,  a  Hermitian structure $(J^{~M}_{N},g_{MN})$, i.e. an almost complex structure, and a positive definite Riemannian metric tensor $g_{MN}$ satisfying the following conditions \cite{nakahara}
\begin{eqnarray}\label{1}
J^{~N}_{R}J^{~M}_{N}=-\delta^{M}_{R},
 \end{eqnarray}
 \begin{eqnarray}\label{2}
g_{MN}J^{~M}_{R}J^{~N}_{S}=g_{RS}.
 \end{eqnarray}
 Then, from the above equations, we have the K\"{a}hler two-form
 \begin{eqnarray}\label{3}
\Omega_{MN}=g_{NR}J^{~R}_{M}=-\Omega_{NM}.
 \end{eqnarray}
The Nijenhuis tensor of an almost Hermitian manifold $(M; J; g)$ is defined as follows
  \begin{eqnarray}\label{6}
 N_{J}(X,Y)=(\nabla_{JX}J)Y-(\nabla_{JY}J)X+J((\nabla_{Y}J)X-(\nabla_{X}J)Y),~~~~~ \forall X,Y \epsilon ~\cal{X}(M)
  \end{eqnarray}
where $\chi(M)$ is the space of vector fields on $M$. Now, if an almost Hermitian structure satisfies the following conditions
\begin{subequations}\label{4}
\begin{eqnarray}
 \nabla_{M}J^{~N}_{R}+\nabla_{R}J^{~N}_{M}=0, \\
 \nabla_{M}J^{~M}_{R}=0,
\end{eqnarray}
\end{subequations}
where the $(5b) $ is the weak form of $(5a)$ and $\nabla$ denotes the operator of covariant derivative with respect to the Riemannian connection, then the manifold is called a\textit{ nearly K\"{a}hler manifold} (or Tachibana space, or K-space) \cite{gary2}. For nearly K\"{a}hler manifold $M$, the Nijenhus tensor does not vanish and we have \cite{yano}
  \begin{eqnarray}\label{7}
 N_{J}(X,Y)=4J(\nabla_{X}J)Y,~~~~~ \forall X,Y \epsilon ~\cal{X}(M).
  \end{eqnarray}
Furthermore, the lowest dimension of the strict nearly K\"{a}hler manifolds is 6 dimensions \cite{yano}. Let $R^{R}_{~SMN}$ and $R_{MN}=g_{RS}R^{R}_{~MSN}$ and $R$ be the \textit{Riemannian curvature tensor}, Ricci tensor and scalar curvature respectively, and $R^{*}_{MN}$ be the Hermitian Ricci tensor which is defined as follows \cite{yano}
 \begin{eqnarray}\label{9}
R^{*}_{MN}\equiv-\frac{1}{2}R_{MK RS}J^{~K}_{N}J^{RS}.
 \end{eqnarray}
 Then, for a nearly K\"{a}hler manifold two versions of Ricci tenors are related by \cite{yano}
 \begin{eqnarray}\label{10}
S_{MN}\equiv R_{MN}-R^{*}_{MN}=-(\nabla_{M}J^{~R}_{S})(\nabla_{N}J^{~S}_{R}),
 \end{eqnarray}
 and
  \begin{eqnarray}\label{11}
S=R-R^{*}=-(\nabla_{M}J^{~R}_{S})(\nabla^{M}J^{~S}_{R})=Constant>0.
 \end{eqnarray}
 In a K-space we have the Bianchi identity for Hermitian Ricci tensor as  $\frac{1}{2}\nabla_{M}R^{*}=\nabla^{N}R^{*}_{NM}$, and by using  \eqref{4} we have \cite{yano}
  \begin{eqnarray}\label{12}
\nabla^{N}(R_{NM}-R^{*}_{NM})=\frac{1}{2}\nabla_{M}(R-R^{*})=0.
 \end{eqnarray}

\section{Curvature-like tensor}
In this section, we investigate properties of a curvature-like tensor on a nearly K\"{a}hler manifold.
For arbitrary constants $a$ and $b$, a tensor which includes Riemannian curvature tensor and almost complex structure could be  defined as follows \cite{yano}
\begin{eqnarray}\label{13}
W_{MNRS}\equiv R_{MNRS}-a(g_{MS}S_{NR}-g_{NS}S_{MR}+g_{NR}S_{MS}-g_{MR}S_{NS})+b(R-R^{*})(g_{MS}g_{NR}-g_{NS}g_{MR}).
 \end{eqnarray}
 There exist real numbers $a$ and $b$ if and only if $d\geq 6$ \cite{yano}.
 $W_{MNRS}$ tensor satisfies the following symmetry properties
 \begin{eqnarray}\label{14}
W_{KLMN}+W_{KMNL}+W_{KNLM}=0,
\end{eqnarray}
\begin{eqnarray}\label{15}
W_{MNRS}=-W_{NMRS}=-W_{MNSR}=W_{RSMN},
\end{eqnarray}
which is then called  a curvature-like tensor \cite{c-L}.

 Because $W_{MNRS}$ tensor has similar terms as those of Weyl tensor, it is appealing
to check the Weyl tensor proprieties for $W_{MNRS}$.  The Weyl tensor is
always invariant under conformal transformation of metric, i.e $\bar{g}=e^{2\sigma}g$; some of useful transformation properties of this tensor are listed in appendix A. The tensor $W_{MNRS}$ is conformal invariant if the arbitrary constants $a$ and $b$ are fixed as $a=-\frac{1}{d-4}$ and $b=-\frac{1}{(d-2)(d-4)}$, and just with these fixed constants $a$ and $b$ the $trW$ will be vanished in conformal flat K-spaces.
However, unlike Weyl tensor whose trace is always zero, $W_{MNRS}$ tensor has non-vanishing trace; in other words for contracted pair of indices in $W_{RMLN}$ we obtain
  \begin{eqnarray}\label{22}
W_{MN}=W^{L}_{~MLN}=R_{MN}-a(2-d)S_{MN}+((1-d)b+a)(R-R^{*})g_{MN}.
 \end{eqnarray}
 Now, contracting \eqref{22} with $g^{MN}$ gives the scalar $W$ as follows
    \begin{eqnarray}\label{23}
W=g^{MN}W_{MN}=R-(d-1)(2a-db)(\nabla_{R}J^{~M}_{N})(\nabla^{R}J^{~N}_{M}).
 \end{eqnarray}
  Therefore, $W_{RMLN}$ tensor cannot be regarded as a Weyl-like tensor, and it suffices to consider it as a curvature-like tensor. On the other hand, being a curvature-like tensor only, $W_{MNRS}$ tensor is not  necessarily  invariant under conformal transformation of the metric; hence, there is no obvious condition for fixing arbitrary constants $a$ and $b$, and they will remain arbitrary from the curvature-like tensor point of view.
\section{Gravity and induced matter}
The main motivation for the present work is to  realize the possible physical signification of the almost complex structure as a geometrical structure carried by some manifolds. The almost complex structure has already been
interpreted as electromagnetic matter field \cite{cfkaluzaklein}. Here, similar
to the gravitational Einstein-Hilbert action where the Ricci scalar $R$ is constructed by the Riemann curvature tensor, we will try to construct a gravitational action by using the scalar curvature \eqref{23} constructed by the curvature-like tensor \eqref{13} which includes the almost complex structure and possesses the symmetry properties of the curvature tensor. One such familiar candidate is the curvature-like tensor  $W_{RMLN}$ of \eqref{13} which  looks like
the Weyl 
 tensor  written in terms of $S_{MN}$ tensor \eqref{10} instead of Ricci tensor $R_{MN}$. Beside this tensor, there are some other known curvature-like tensors containing $J$-dependent terms, for example the holomorphic
 curvature tensor introduced in \cite{hct}. However, the motivation for choosing a curvature-like 
tensor $W_{RMLN}$,  in  comparison with other tensors mentioned above, is that the tensor $W_{RMLN}$ is potentially  decomposable to the Einstein-Hilbert action plus a  
matter type action, as is shown in the following. Furthermore, as it will 
 be seen in the following and in appendix $B$, the trace of this tensor has 
 a term which could be recognized as strength tensor of the almost complex 
 structure (or equivalently of K\"{a}hler two form $\Omega_{MN}$) \footnote{The holomorphic
  curvature tensor  has been used for constructing an invariant tensor under conformal 
  transformation of metric, named generalized Bochner curvature tensor which is related to the Weyl tensor \cite{WB}. Similarly, one could consider a tensor based on the curvature-like tensor $W_{RMLN}$ with 
  $U_{RMLN}\equiv~^{*}O_{RM}^{~~KP}W_{KPLN}=\frac{1}{2}(\delta_{R}^{~K}\delta_{M}^{~P}+J_{R}^{~K}J_{M}^{~P})W_{KPLN}$, which beside 
  being a curvature-like tensor, could be conformal invariant with fixed arbitrary constants $a$ and $b$ and particularly has 
  vanishing trace (i.e. $U=g^{RL}g^{MN}=0$) and so it could be recognized as a type of generalization of Weyl tensor. From this point of view, the $W_{RMLN}$ has been written in such a way that leads to a Weyl-like tensor.}.

 The scalar \eqref{23} contains Ricci scalar $R$ and a term carrying covariant derivatives of the almost complex structure. We start with a general situation where there is no emphasis on $J^{~M}_{N}$ being a nearly K\"{a}hler structure, and \textit{the only requirement is non-K\"{a}hlerity}, $\nabla J\neq 0$. Then, the action with this scalar in {\it even} $d$-dimensions ($d\geq 6$) is given by\footnote{We have used the natural system of units with $G=c=1$ and the dimension dependent factor of the cosmological constant term has been chosen to have $R_{MN}=\Lambda g_{MN}$ in vacuum.}
  \begin{eqnarray}\label{24}
S&=&\frac{1}{16\pi}\int d^{d}x\sqrt{-g} (W-(d-2)\Lambda)\nonumber\\
&=&\frac{1}{16\pi}\int d^{d}x\sqrt{-g}(R-(d-2)\Lambda)-\dfrac{(d-1)(2a-db)}{16\pi}\int d^{d}x\sqrt{-g}(\nabla^{K}J^{~M}_{N})(\nabla_{K}J^{~N}_{M})\nonumber\\
&\equiv&\frac{1}{16\pi} S_{EH}+S_{M}(J).
 \end{eqnarray}

In this way, the almost complex structure $J$ as a part of geometrical data enters in the action via the scalar of the curvature-like 
tensor. Briefly, the almost complex manifold is equipped with  the metric $g$ and the almost complex structure  $J$ which are two geometrical structures on the 
manifold and the above action encompasses both of them where the $d$-dimensional Einstein-Hilbert term is decomposed totally from the $d$-dimensional $J$ terms \footnote{The coupling of $J$ terms with metric is introduced from gravitational point of view.}. Now, 
one may consider the second term in \eqref{24} as a type of matter action and  obtain a gravitational model which contains the Einstein-Hilbert action and a matter term which is induced by the almost complex 
structure in a geometrical way. \textit{The key point
here is that the obtained matter action is not considered as an additional action to the Einstein-Hilbert action, rather it is obtained in a straightforward way from a pure geometric based action including a curvature-like 
tensor $W_{RMLN}$}.

Conservation of energy momentum tensor derived from the above action requires the almost complex structure  to be nearly K\"{a}hlerian type. However, the equations of motion of $J$ would kill the energy momentum tensor in the nearly K\"{a}hler case. The strategy for solving this inconsistency is to consider the geometry as nearly K\"{a}hler type and add the nearly K\"{a}hler condition $(5)$ with Lagrange multiplier to the action \eqref{24}.  Since this energy-momentum tensor will be derived from the structure of nearly K\"{a}hler manifold, we shall call it as the {\it geometrically induced matter}. Hence, we reconsider the action \eqref{24} within nearly K\"{a}hler structure which guarantees the conservation of energy momentum tensor, as follows
\begin{eqnarray}\label{model}
S=\frac{1}{16\pi}S_{EH}+S_{M}(J)+\int d^{d}x\sqrt{-g}\lambda^{M}\nabla_{N}J_{M}^{~~N}.
   \end{eqnarray} 
Variations with respect to $\lambda^{M}$ and $J_{M}^{~N}$ give the equations of motion as follows
 \begin{eqnarray}\label{l}
  \nabla_{N}J_{M}^{~~N}=0,
   \end{eqnarray}
   \begin{eqnarray}\label{25}
   -2(d-1)(2a-db)\nabla^{R}\nabla_{R}J_{M}^{~N}+\nabla_{M}\lambda^{N}=0.
   \end{eqnarray}
   Employing the equation of motion of $J_{M}^{~N}$ \eqref{25} for eliminating the Lagrange multipliers $\lambda^{M}$ gives the following extremized action after integrating by parts
   \begin{eqnarray}\label{250} 
   S=\frac{1}{16\pi}\int d^{d}x\sqrt{-g}(R-(d-2)\Lambda)-\dfrac{(d-1)(2a-db)}{16\pi}\int d^{d}x\sqrt{-g}(\nabla^{K}J^{~M}_{N})(\nabla_{K}J^{~N}_{M}).
   \end{eqnarray}
As is mentioned in detail in appendix $B$, the second term in the action has a form of $H_{MNK}^{(nk)}H^{(nk)~MNK}$, where $H_{MNK}^{(nk)}$ is the field strength associated with the nearly K\"{a}hler
two form $\Omega_{MN}=g_{NK}J_{M}^{~K}$ given in \eqref{90}.  
 
 Using the general formula for the symmetric energy-momentum tensor 
 \begin{eqnarray}\label{26}
 T_{MN}=\frac{\delta(\sqrt{-g} \cal L_{M})}{\delta g^{MN}}=g_{MN}{\cal L_{M}}-2\frac{\delta{\cal L_{M}}}{\delta g^{MN}},
 \end{eqnarray}
 and varying matter Lagrangian with respect to the metric leads to symmetric energy-momentum tensor as follows
 \begin{eqnarray}\label{27}
 T_{MN}=\dfrac{(d-1)(2a-db)}{16\pi}[-2(\nabla_{M}J^{~K}_{L})(\nabla_{N}J^{~L}_{K})-4(\nabla_{L}J_{KN})(\nabla^{L}J^{~K}_{M})\nonumber\\
 +g_{MN}(\nabla_{P}J^{~K}_{L})(\nabla^{P}J^{~L}_{K})],
  \end{eqnarray}
 where we have used $J^{2}=-1$. The non-zero contribution of the Lagrange multiplier term in \eqref{25} prevents the energy momentum tensor from being vanished due to the equation of motion of $J$. In general, varying the action \eqref{250} with respect to the metric results in Einstein equations as follows
  \begin{eqnarray}\label{28}
R_{MN}-\frac{1}{2}R~g_{MN}+\frac{d-2}{2}\Lambda ~g_{MN}=8\pi T_{MN},
 \end{eqnarray}
 so that the almost complex structure $J$ as a geometric structure on manifold appears in the right hand side of Einstein equation as an induced energy-momentum tensor. 
Now, by considering the covariant derivative of \eqref{27} to investigate the conservation of energy-momentum tensor and using \eqref{10}, \eqref{11}, we have
   \begin{eqnarray}\label{29}
\nabla^{M}T_{MN}=\dfrac{(d-1)(2a-db)}{16\pi}[-2(\nabla^{M}(R_{MN}-R^{*}_{MN})-\dfrac{1}{2}\nabla_{N}(R-R^{*}))-4\nabla^{M}(\nabla_{L}J_{KN})(\nabla^{L}J^{~K}_{M})],
 \end{eqnarray}
 the above equation is covariantly constant if both nearly K\"{a}hler properties in \eqref{4} be taken into account with using the Bianchi-like identity \eqref{12}. In fact, we have the weak nearly K\"{a}hler condition $(5b)$ as the equation of motion \eqref{l}, when the general and stronger condition of nearly K\"{a}hler structure $(5a)$, is required by the conservation of energy-momentum tensor. 
 Note that the second term could be rewritten in terms of $\nabla^{M}(R_{MN}-R^{*}_{MN})$ by using $(5a)$.
 In the last calculations we have used the properties of curvature tensor in nearly K\"{a}hler manifolds
  $R_{MNRS}=J^{~K}_{M}J^{~L}_{N}J^{~P}_{R}J^{~Q}_{S}R_{KLPQ}=J^{~K}_{R}J^{~L}_{S}R_{MNKL}-(\nabla_{M}J^{~L}_{N})(\nabla_{R}J_{SL})$
   and the relation $S^{NM}J_{N}^{~~K}=-S^{NK}J_{N}^{~~M}$ \cite{yano}.
So, the matter in the action is minimally coupled to gravity if the matter field $J$ be of nearly K\"{a}hler type, then 
  \begin{eqnarray}
  \nabla^{M}T_{MN}=0.
   \end{eqnarray}
 
   Now, employing the nearly K\"{a}hlerian properties \eqref{4} results in
the energy-momentum tensor as follows
\begin{eqnarray}\label{299}
T_{MN}=\dfrac{(d-1)(2a-db)}{16\pi}[-6(\nabla_{M}J^{~K}_{L})(\nabla_{N}J^{~L}_{K})
+g_{MN}(\nabla_{P}J^{~K}_{L})(\nabla^{P}J^{~L}_{K})],
 \end{eqnarray}   
 where its trace is given by
    \begin{eqnarray}\label{30}
T&=&g^{MN}T_{MN}\nonumber\\
&=&\dfrac{(d-1)(2a-db)(d-6)}{16\pi}(\nabla_{P}J^{~K}_{L})(\nabla^{P}J^{~L}_{K}).
 \end{eqnarray}
Obviously, the trace of energy momentum tensor does not vanish. In fact, $T_{MN}$ is traceless for dilation invariant scalar field theories, so conformal invariance has been broken.
Moreover, under general coordinate transformation
 \begin{eqnarray}
 x'^{M}=x^{M}+\epsilon~ \xi^{M}(x),~~~~~~\epsilon\rightarrow 0,
 \end{eqnarray}
 where the metric and K\"{a}hler two-form transform as 
\begin{eqnarray}
g'_{MN}-g_{MN}&=&\epsilon\partial _{M}\xi ^{K}g_{KN}+\epsilon\partial _{N}\xi ^{K}g_{MK}+\epsilon\xi ^{K}\partial _{K}g_{MN},\nonumber\\
\Omega'_{MN}-\Omega_{MN}&=&\epsilon\partial _{M}\xi ^{K}\Omega_{KN}+\epsilon\partial _{N}\xi ^{K}\Omega_{MK}+\epsilon\xi ^{K}\partial _{K}\Omega_{MN},
\end{eqnarray}
a direct calculation by using a relation between covariant derivative of $J$, Ricci and Hermitian Ricci tensor on nearly K\"{a}hler manifolds as \cite{yano}
\begin{eqnarray}\label{330}
\nabla_{M}\nabla^{M}J_{K}^{~~N}=J^{NM}(R_{KM}-R^{*}_{KM}),
\end{eqnarray}
reveals that the action \eqref{model} is invariant under general coordinate transformation if we have $\partial.\xi=0$.

\section{Example}
In this section, we introduce a particular 10-dimensional nearly K\"{a}hler manifold on which we would like to describe our gravitational model \eqref{model}. It is a theorem that a nearly K\"{a}hler manifold $M$ can be decomposed as direct product $M=M^{k}\times M^{s}$, where $M^{k}$ is a K\"{a}hler and $M^{s}$ is a strictly nearly K\"{a}hler manifold \cite{gary2}. The known 6-dimensional examples of nearly K\"{a}hler manifolds are $SU(3)/U(1)\times U(1)$, $G_{2}/SU(3)$, $SP(2)/SU(2)\times U(1)$ and $S^3\times S^3$ \cite{Butruille}.
Particularly, we are interested in group manifolds, so we will focus on $S^3\times S^3$ which is the only group manifold example of known nearly K\"{a}hler manifolds, up to now. Then, with a 4-dimensional K\"{a}hler manifold $M_{4}$ (i.e. ${\nabla}J=0$), $M_{4}\times S^3\times S^3$ will be a 10-dimensional nearly K\"{a}hler manifold on which we can consider a metric ansatz of the form
 \begin{eqnarray}\label{31}
dS^{2}_{10}=g_{MN}dx^{M}dx^{N}
=g_{\mu\nu}(x^{\rho})dx^{\mu}dx^{\nu}+g_{\hat{\mu}\hat{\nu}}(x^{\hat{\rho}})dx^{\hat{\mu}}dx^{\hat{\nu}},
 \end{eqnarray}
where $g_{\mu\nu}(x^{\rho})$ is a 4-dimensional space-time metric and $g_{\hat{\mu}\hat{\nu}}(x^{\hat{\rho}})$ denotes the 6-dimensional metric. The indices $M, N, ...$ run over the whole 10 dimensions, the indices $\mu, \nu, \rho, ...$ run over $ 0, 1, 2, 3$  labeling 4-dimensional space-time, the indices $\hat{\mu}, \hat{\nu},\hat{\rho}$
run over $4,5,...9$ labeling 6-dimensional compact nearly K\"{a}hler manifold.
It is more convenient to apply our formalism in non-coordinate basis \cite{nakahara}
and for this reason we will prefer a group manifold to workout as an example\footnote{In the coordinate basis $T_{p}M$ is spanned by $\{e_{M}\}=\{\frac{\partial}{\partial x^{M}}\}$, where in non-coordinate basis there is an alternative choice for the basis as $\{e_{A}\}$. These two basis are related to each other by vielbines $e_{M}^{~~A}$  and we have $[e_{A},e_{A}]=f_{AB}^{~~C}e_{C}$.}.

Consider a 10-dimansional group manifold, $R\times B\times SU(2)\times SU(2)$, where $R$ is 1-dimensional Abelian Lie group whose coordinate will be regarded as time variable, and $B$ is a 3-dimensional real Lie group (Bianchi Lie group) \cite{Landau}. Basis of $B$ are labeled by $i, j, k, ...$,  and  indices $\hat{a}, \hat{b}, \hat{c},..$ run over whole Lie algebra $SU(2)\times SU(2)$ where indices $A, B, C, ...$ will label 10-dimensional manifold.
We consider a non-coordinate basis for the 4-dimensional part of manifold where the vielbins $e_{\mu}^{~~a}(x)$ depend only on the space coordinates, and the non-coordinate metric $g_{ab}(t)$ as the variable of $R$ Lie group depends only on time $t$. In this way, factorizing 4-dimensional space-time metric in a synchronous frame gives \cite{Landau,mr}
 \begin{eqnarray}\label{34}
 g_{\mu\nu}dx^{\mu}dx^{\nu}&=&e_{\mu}^{~~a}(x) g_{ab}(t)e_{\nu}^{~~b}(x)dx^{\mu}dx^{\nu}\nonumber\\
&=&-g_{00}(t)dt^{2}+e_{\alpha}^{~~i}(x) g_{ij}(t)e_{\beta}^{~~j}(x)dx^{\alpha}dx^{\beta},
  \end{eqnarray}
where $ {X_{i}} $ and $ {x_{i}} $ indicate generators and coordinates of $B$ Lie group, respectively. Then, for 6-dimensional space we set
\begin{eqnarray}\label{35}
 g_{\hat{\mu}\hat{\nu}}=e_{\hat{\mu}}^{~~\hat{a}}(x_{\hat{a}}) e_{\hat{\nu}}^{~~\hat{b}}(x_{\hat{a}}) g_{\hat{a}\hat{b}},
\end{eqnarray}
where $ {{X_{\hat{a}}}} $ and $ {x_{\hat{a}}} $ are the generators and coordinates of $SU(2)\times SU(2)$ Lie group, respectively.
In this non-coordinate basis for 10-dimensional manifold we have the following relation for Ricci tensor
  \begin{eqnarray}\label{36}
   R_{MN}=e_{M}^{~~a} e_{N}^{~~b} R^{(4)}_{ab}(t)+e_{M}^{~~\hat{a}} e_{N}^{~~\hat{b}} R^{(6)}_{\hat{a}\hat{b}}.
  \end{eqnarray}
   Note that, as in 6-dimensional Einstein-Yang-Mills theory \cite{ranjbar}, Poincare invariance implies
   \begin{eqnarray}\label{360}
    R_{a\hat{b}}=0,~~~~~ \Omega_{a\hat{b}}=0,
     \end{eqnarray}
hence $J_{a}^{~\hat{b}}=0$. The expression of Ricci tensor in non-coordinate basis in terms of structure constants and time derivatives is given in the
appendix $C$ in \eqref{38}.
      
Returning to the Einstein equations of motion \eqref{28}, the metric and $J$ are not completely arbitrary and should be of nearly K\"{a}hler type. Hence, the first essential step toward solving the equations of motion is to identify a nearly K\"{a}hler metric and almost complex structure by solving the equations of $(5a)$ and $(5b)$ along with \eqref{1} and \eqref{2}. As mentioned in the previous section,  the nearly K\"{a}hler condition imposes an extra identity on $J$ which is consistent with the metric and  the conservation of energy momentum tensor. In the appendix $C$, our new method of calculating nearly K\"{a}hler structure in the non-coordinate basis is explained in detail, where for a particular example of $R\times II\times SU(2)\times SU(2)$ \footnote{
Note that $II$ denotes the Bianchi type Lie group which along with $R$ Lie group is capable of having a class of K\"{a}hler 
structure given in \eqref{55} and \eqref{555}.}  the  metric and complex structure have been obtained in \eqref{44}, \eqref{444} and \eqref{55}, \eqref{555}. As 
mentioned above, the $SU(2)\times SU(2)$ part of 10 dimensional group manifold is strictly nearly  K\"{a}hler and the $R\times 
II$ part of it is a K\"{a}hler manifold, so  they altogether construct a strictly nearly  K\"{a}hler manifold. The final nearly  K\"{a}hler 
structure (metric and complex structure) which contains an arbitrary function of time $F(t)$ up to four arbitrary constants $c_{1}, c_{2}, c_{3}$ and $\xi$ is given by
\begin{eqnarray}\label{111}
dS^{2}_{10}&=& g_{ab}e^{a}\otimes  e^{b}+g_{\hat{a}\hat{b}}e^{\hat{a}}\otimes  e^{\hat{b}}\nonumber\\
 &=&(-{\frac {\rm d}{{\rm d}t}}F \left( t \right)e^{1}\otimes  e^{1}-{{\it c_{3}}}^{2}{\frac {\rm d}{{\rm d}t}}F \left( t \right) e^{2}\otimes  e^{2}+{\it c_{1}}+{\it c_{2}}F \left( t \right) e^{3}\otimes  e^{3}+{\frac { \left( {\it c_{1}}+F \left( t \right) {\it c_{2}} \right) {{\it c_{3}}
}^{2}}{{{\it c_{2}}}^{2}}}e^{4}\otimes  e^{4})\nonumber\\
&-&( 2\xi(e^{5}\otimes  e^{5}+e^{6}\otimes  e^{6}+e^{7}\otimes  e^{7}+e^{8}\otimes  e^{8}+e^{9}\otimes  e^{9}+ e^{10}\otimes  e^{10}\nonumber\\
&-&\frac{1}{2}(e^{5}\otimes  e^{8}+e^{6}\otimes  e^{9}+e^{7}\otimes  e^{10}))),
     \end{eqnarray}
\begin{eqnarray}\label{112}
\Omega&=&\frac{1}{2}(\Omega_{ab}e^{a}\wedge e^{b}+\Omega_{\hat{a}\hat{b}}e^{\hat{a}}\wedge  e^{\hat{b}})\nonumber\\
&=&\frac{1}{2}(-{\it c_{3}}\,{\frac {\rm d}{{\rm d}t}}F \left( t \right)~~ e^{1}\wedge  e^{2}+{\frac {{\it c_{3}}\, \left( {\it c_{1}}+F \left( t \right) {\it c_{2}}
 \right) }{{\it c_{2}}}}~~e^{3}\wedge  e^{4}+\sqrt {3}\xi\ (e^{5}\wedge  e^{8}+e^{6}\wedge  e^{9}+e^{7}\wedge  e^{10})
).
     \end{eqnarray}
The metric and almost complex structure present nearly K\"{a}hler structure with some arbitrariness which could be determined by  solving the Einstein equations. The metric and K\"{a}hler two form in the equations are in the non-coordinate basis and explicit forms of $g_{MN}$ and $\Omega_{MN}$ may be obtained by multiplication of the vielbien given in \eqref{vs} and \eqref{vII}. 

Now, with the above nearly K\"{a}hler metric and complex structure we may solve Einstein equations to fix the function $F(t)$, and take the advantageous of the arbitrary constants to construct the correct signature of metric. Decomposing Einstein equations \eqref{28} in 6 extra dimensions and 4-dimensional space-time gives
  \begin{eqnarray}\label{32}
  R_{\hat{\mu}\hat{\nu}}-\frac{1}{2}R^{(10)} g_{\hat{\mu}\hat{\nu}}+4\Lambda g_{\hat{\mu}\hat{\nu}}=\dfrac{(18 a-90 b)}{2}(-6(\nabla_{\hat{\mu}}J^{~\hat{\rho}}_{\hat{\sigma}})(\nabla_{\hat{\nu}}J^{~\hat{\sigma}}_{\hat{\rho}})+g_{\hat{\mu}\hat{\nu}}(\nabla_{\hat{\lambda}}J^{~\hat{\rho}}_{\hat{\sigma}})(\nabla^{\hat{\lambda}}J^{~\hat{\sigma}}_{\hat{\rho}})),
   \end{eqnarray}
    \begin{eqnarray}\label{33}
    R_{\mu\nu}-\frac{1}{2}R^{(10)} g_{\mu\nu}+4\Lambda g_{\mu\nu}=\frac{(18a-90b)}{2
    }g_{\mu\nu}(\nabla_{\hat{\lambda}}J^{~\hat{\rho}}_{\hat{\sigma}})(\nabla^{\hat{\lambda}}J^{~\hat{\sigma}}_{\hat{\rho}}),
     \end{eqnarray}
where, noting \eqref{360}, the Ricci scalar of 10-dimensional manifold is the sum of Ricci scalar of 4$d$ and 6$d$ parts of manifold, i.e. $R^{(10)}=R^{(4)}+R^{(6)}$. On K\"{a}hler part of the manifold, $M_{4}$, the first term of energy momentum tensor \eqref{27} vanishes, but a non-zero contribution in the second term will be inherited from the 6-dimensional nearly K\"{a}hler manifold $S^3\times S^3$.
Solving the above Einstein equations over the metric \eqref{111} and the
complex structure \eqref{112}   by using \eqref{3}, and choosing arbitrary constants as $c_{1}=1$, $c_{2}=-1$, and
$c_{3}=1$, gives the function $F(t)$ and the cosmological constant, respectively
as follows
 \begin{eqnarray}\label{F(t)}
 F(t)={\frac { \left( -24\,\Lambda\,\xi+280\,a-1400\,b-5 \right) t+ \left( -
 24\,\Lambda+9 \right) \xi+280\,a-1400\,b-5}{ \left( t+1 \right) 
  \left( -24\,\Lambda\,\xi+280\,a-1400\,b-5 \right) }}
 ,
 \end{eqnarray}
  \begin{eqnarray}
 \Lambda=\dfrac{1}{18}\,{\frac {264~a-1320~b-5}{\xi}}.
 \end{eqnarray}
  Consequently, by a redefinition of time as $t={{\rm e}^{\tau}}-1$, we obtain the 4-dimensional part of metric \eqref{34} and the almost complex structure of 4-dimensional space-time, respectively as 
 \begin{eqnarray}
  g_{\mu\nu}={\frac {-27\xi}{216 a-1080 b+5}}(-d\tau^{2}+{{\rm e}^{-\tau}}dx_{1}^{2}-x_{3}{{\rm e}^{-2\tau}}dx_{1}dx_{2}+ (-x_{3}^{2}{{\rm e}^{-2\tau}}+{{\rm e}^{-\tau}})dx_{2}^{2}+ {{\rm e}^{-\tau}} dx_{3}^{2}),
  \end{eqnarray}
 \begin{eqnarray}
 J_{\mu}^{~\nu}=(\dfrac{27\xi}{216 a +1080 b +5})^{2}\left[ \begin {array}{cccc} 0&{{\rm e}^{-3\,\tau}} \left( {x_{3}}^{2}
 +1 \right) &-{{\rm e}^{-3\,\tau}}x_{3}\, \left( -{x_{3}}^{2}+{{\rm e}^
 {\tau}}-1 \right) &0\\ \noalign{\medskip}-{{\rm e}^{-\tau}}&0&0&0
 \\ \noalign{\medskip}-{{\rm e}^{-\tau}}x_{3}&0&0&-{{\rm e}^{-2\,\tau}}
 \\ \noalign{\medskip}0&-{{\rm e}^{-3\,\tau}}x_{3}&{{\rm e}^{-3\,\tau}}
  \left( -{x_{3}}^{2}+{{\rm e}^{\tau}} \right) &0\end {array} \right]_.
   \end{eqnarray}

 Also, it turns out that the 4-dimensional part of the energy-momentum tensor has a direct relation with space-time metric as 
  \begin{eqnarray}\label{CC}
  T_{\mu\nu}={\frac {1260\,a-6300\,b}{27\,\xi}}g_{\mu\nu},
  \end{eqnarray}
i.e. the 4-dimensional induced matter obtained in this example is in the form of a {\it cosmological constant} which depends explicitly on the parameters $a, b$ and $\xi$, among which $\xi$ is an element of $6d$ part of the manifold. In this regard, it is appealing to  discuss about the cosmological constant and its known problems in the context of the present model.
\subsection{Cosmological constant and the fine-tuning  problem}

According to the present observations, the experimental upper bound on the current value of the cosmological constant is extremely small. Moreover, it is usually assumed that an effective cosmological constant describes the energy density of the vacuum $<\rho_{vac}>$. Actually, it is commonly
believed that the vacuum energy density $<\rho_{vac}>$ contains the quantum field theory contributions to the effective cosmological constant
\begin{equation}\label{1'}
\Lambda_{eff}=\Lambda+ \kappa <\rho_{vac}>,
\end{equation}
where $\Lambda$ is a bare cosmological constant.
On the other hand, the calculations show that the quantum field theory contributions affect enormously the value of effective cosmological constant as \cite{carroll}
\begin{equation}\label{3'}
<\rho_{vac}> \sim
M_{EW}^4  \sim 10^{47} {\rm ~erg/cm}^3\ ,
\end{equation} 
for electroweak cut off
\begin{equation}\label{4'}
<\rho_{vac}> \sim
M_{QCD}^4  \sim  10^{36} {\rm ~erg/cm}^3\ ,
\end{equation}
for QCD cut off 
\begin{equation}\label{5'}
<\rho_{vac}> \sim
M_{GUT}^4  \sim 10^{102} {\rm ~erg/cm}^3\ ,
\end{equation} 
for GUT cut off, and 
\begin{equation}\label{6'}
<\rho_{vac}> \sim
M_P^4  \sim  10^{110} {\rm ~erg/cm}^3\ ,
\end{equation} 
for Planck cut off. General relativity as a classical theory is applied on the scales larger than the Planck scale so that one may reasonably expect that theoretically the Einstein equation in 4-dimensional space-time is certainly valid for electroweak, QCD, GUT scales and is almost valid for Planck scale (in $G=1$ units) as
\begin{equation}\label{7'}
R_{\mu\nu} - {1\over 2}R^{(4)}g_{\mu\nu}+ \Lambda_{eff}\, g_{\mu\nu}
= 8\pi T_{\mu\nu}\,, 
\end{equation}
where $\Lambda_{eff}$ is the effective cosmological constant with contributions
coming from electroweak, QCD, GUT and even Planck scales. However, the current
observational considerations requires the following Einstein equation
\begin{equation}\label{8}
R_{\mu\nu} - {1\over 2}R^{(4)}g_{\mu\nu} 
+ \Lambda_{obs} \,g_{\mu\nu}
= 8\pi T_{\mu\nu}\,, 
\end{equation}
where $\Lambda_{obs}$ is the observed cosmological constant corresponding
to an energy density with the order of magnitude $10^{-10} {\rm ~erg/cm}^3$. The fact that $\Lambda_{obs} \sim 10^{-120} \Lambda_{eff}$ is the
well-known cosmological constant problem \cite{Weinberg}.
{Many approaches have been introduced to solve this challenging problem with
no full success. Hence, some people have been interested in finding an alleviation
to this problem by resorting to a fine-tuning mechanism
which at least can provide us with a small observed value for the cosmological
constant.} In the above example, we found the induced 4-dimensional energy-momentum tensor \eqref{CC} in the form of a cosmological constant. Hence, the 4-dimensional Einstein equation \eqref{33} can be written as
\begin{eqnarray}\label{33'}
    R_{\mu\nu}-\frac{1}{2}R^{(4)} g_{\mu\nu}+4\Lambda g_{\mu\nu}=(
    {\frac {1260\,a-6300\,b}{27\,\xi}}+\frac{1}{2}R^{(6)})g_{\mu\nu},
     \end{eqnarray}
or effectively as 
\begin{eqnarray}\label{33''}
    R_{\mu\nu}-\frac{1}{2}R^{(4)} g_{\mu\nu}+\Lambda_{obs}\, g_{\mu\nu}=0,
     \end{eqnarray}
where
\begin{equation}\label{obs}
\Lambda_{obs}=\dfrac{2}{9}{\frac {264\,a-1320\,b-5}{\xi}}-{\frac {1260\,a-6300\,b}{27\,\xi}}-\frac{1}{2}R^{(6)}.
\end{equation}
Eq.\eqref{33''} shows an Einstein equation where the induced matter \eqref{CC}
is removed in favour of an effective cosmological constant $\Lambda_{obs}$ which is capable of being fine-tuned to a very small or vanishing value, in agreement with observations, by the free parameters $a, b$ and $\xi$. Actually, the parameter $\xi$ depends on the $6d$ part of manifold through \eqref{44} and \eqref{38} and is related to the Ricci scalar of the $6d$ part as $R^{(6)}=\frac{-5}{3 \xi}$.
Setting $\Lambda_{obs}\simeq0$ gives 
$$
a\simeq{5\,b+{\frac {5}{1896}}},
$$
or equivalently the factor $(18a - 90b)$  in Lagrangian will be fixed as $(\simeq){\dfrac{15}{316}}$.

 \section{Candidate for dark energy }
 
 Recent cosmological observations {\cite{c1}}, WMAP {\cite{c2}}, SDSS {\cite{c3}} and X-ray {\cite{c4}} indicate that our universe is really experiencing
an accelerated expansion. These observations confirm also that the
universe is spatially flat, and consists of about $70 \%$ dark
energy with negative pressure, $30\%$ dust matter (cold dark
matter plus baryons), and some negligible radiation. 

To explain the nature of dark energy and the origin of cosmic acceleration, many theories and models have been proposed. The simplest candidate for the dark energy is considered as a tiny positive cosmological constant. An alternative proposal to explain the dark energy is the dynamical dark energy scenario where the effective dynamical nature of dark energy can originate from various fields, such as canonical scalar field (quintessence) \cite{quint}, phantom field \cite{phant}, or the combination of quintessence and phantom in a unified model named quintom
\cite{quintom}. Another theory has recently been constructed in the light of the holographic principle of quantum gravity which may simultaneously provide a solution to the coincidence problem \cite{holoprin}. 

Let us now investigate if the almost complex structure can play the role
of  dark energy. In this regard, we add a $4d$ baryonic matter term
to the action \eqref{model} as \footnote{Note that as mentioned in introduction,
 the Einstein-Hilbert term, the $\nabla J \nabla J$  term and the matter term do not include terms with higher than \textit{second} derivative of metric.
}
\begin{eqnarray}\label{24'}
S=\frac{1}{16\pi}\int d^{d}x\sqrt{-g}(R-(d-2)\Lambda)+\dfrac{(d-1)(2a-db)}{16\pi}\int d^{d}x\sqrt{-g}(\nabla^{K}J^{~M}_{N})(\nabla_{K}J^{~N}_{M})+\int d^{4}x\sqrt{-g}\, (\textit{L}_M).
 \end{eqnarray}
In $d=10$, it is easy to show that the $4d$ field equations take the following form
\begin{eqnarray}\label{33'''}
    R_{\mu\nu}-\frac{1}{2}R^{(4)} g_{\mu\nu}+4\Lambda g_{\mu\nu}=8 \pi T_{\mu\nu}^{(M)}+\frac{(18a-90b)}{2
    }g_{\mu\nu}(\nabla_{\hat{\lambda}}J^{~\hat{\rho}}_{\hat{\sigma}})(\nabla^{\hat{\lambda}}J^{~\hat{\sigma}}_{\hat{\rho}})+\frac{1}{2}R^{(6)} g_{\mu\nu},
     \end{eqnarray}
where $T_{\mu\nu}^{(M)}$ is the baryonic matter energy-momentum tensor. Now,
we may rewrite \eqref{33'''}
in the following form
\begin{eqnarray}\label{33''''}
    R_{\mu\nu}-\frac{1}{2}R^{(4)} g_{\mu\nu}=8\pi (T_{\mu\nu}^{(M)}+T_{\mu\nu}^{(DE)}),
\end{eqnarray}    
where $T_{\mu\nu}^{(DE)}$ is considered as the dark energy constructed entirely by the
almost complex structure as
\begin{eqnarray}\label{33'''''}   
T_{\mu\nu}^{(DE)}=\frac{(18a-90b)}{2}g_{\mu\nu}(\nabla_{\hat{\lambda}}J^{~\hat{\rho}}_{\hat{\sigma}})
(\nabla^{\hat{\lambda}}J^{~\hat{\sigma}}_{\hat{\rho}})-4\Lambda g_{\mu\nu}+\frac{1}{2}R^{(6)} g_{\mu\nu}.
\end{eqnarray}
For example, in the case of $R\times II \times S^{3}\times S^{3}$ the dark
energy $T_{\mu\nu}^{(DE)}$   
becomes $\Lambda_{obs}$  given by \eqref{obs}. Considering $T_{\mu\nu}^{(M)}$
as the perfect fluid and $g_{\mu\nu}$ as the Friedman-Robertson-Walker
metric, the Einstein equations lead to the following acceleration
equation
\begin{eqnarray}
\frac{\ddot{a}}{a}=-\frac{1}{6}[\rho+3(p-\Lambda_{obs})].
\end{eqnarray}
At present era, where we have a pressureless universe $p=0$, the above equation
indicates that the universe is accelerating provided that the current matter
density $\rho$ is smaller than the dark energy density $3\Lambda_{obs}$ namely $\rho<3\Lambda_{obs}$.
{\it In other words, we have a $\Lambda$CDM model where the dark energy is provided by the almost complex structure}.

At the end of this section it is worth mentioning that according to \eqref{30}, the trace of energy-momentum tensor is proportional to the second term in the action, so the Lagrangian of \eqref{24'} may be regarded as a type of $f(R,T)$ modified theory of gravity which is currently considered as
alternative to dark energy model \cite{FRT}
\begin{eqnarray}
 S=\int d^{d}x\sqrt{-g}(~\frac{1}{16\pi}~f(R,T)+L_{M}).
 \end{eqnarray}
 In a special case of $f(R,T)=R+2f(T)$, our model is equivalent to   $$
 f(T)=\dfrac{8\pi}{(d-6)}T-\frac{(d-2)}{2}\Lambda.
 $$ 
\section{Conclusion}

 In this work, we have investigated a curvature-like tensor on nearly K\"{a}hler manifolds, which beside possessing of the symmetry properties of curvature tensor, carries almost complex structure and may be invariant under conformal transformation of metric. Then, following the idea of including 
  the almost complex structure in action integral, we constructed a gravitational model with the scalar curvature 
  of the curvature-like tensor instead of the Ricci scalar in Einstein-Hilbert action. 
  Therefore, the almost complex structure appeared in the action integral strictly by a geometrical way. Moreover, the corresponding energy-momentum tensor  is interpreted as a dark energy term.  It is remarkable that the conservation of energy-momentum tensor 
  and diffeomorphism invariance in the coordinates coincide with the nearly K\"{a}hlerian properties of the manifold. 
  Furthermore, we have identified a nearly K\"{a}hler complex structure with associated Hermitian metric on a 
  particular example of group manifold $R\times II\times S^{3}\times S^{3}$. The formalism has been developed 
  in non-coordinate basis which greatly simplified the analysis. It turned
  out that the obtained structure on $S^{3}\times S^{3}$ part 
  of manifold is in accordance with the structure which has been found in Ref. \cite{halfflat}. 
  Then, we have solved the Einstein field equations exactly and obtained
  the corresponding $4d$ metric and energy-momentum tensor. The 
energy-momentum tensor as the induced matter appeared in the form of a cosmological
constant. We studied the  cosmological constant and found a solution for
the fine-tuning problem. Moreover, it turned out that the almost complex structure may be considered
as potential candidate for dark energy.
   There are some  interesting open problems which deserve to be investigated
as: {\it What is the two dimensional sigma model whose 4-dimensional effective field theory is identified with our model}. These and some other interesting problems are under our current investigation.
\section*{Acknowledgment}
This research has been supported by Azarbaijan Shahid Madani university by a research fund No. 401.231.
\section*{Appendix A}
In this appendix, we list the behaviors of tensors under conformal transformation of the metric, $\bar{g}=e^{2\sigma}g$.  Riemann curvature tensor transforms as \cite{nakahara}
  \begin{eqnarray}\label{16}
\bar{R}^{S}_{~LMN}=R^{S}_{~LMN}-g_{NL}B_{M}^{~S}+g_{KL}B_{M}^{~K}\delta^{S}_{N}-g_{KL}B_{N}^{~K}\delta^{S}_{M}+g_{ML}B_{N}^{~S},
 \end{eqnarray}
  where
 \begin{eqnarray}\label{17}
B_{M}^{~K}=-\partial_{M}\sigma g^{KL}\partial_{L}\sigma+g^{KL}(\partial_{M}\partial_{L}\sigma-\Gamma^{P}{}_{ML}\partial_{P}\sigma)+\frac{1}{2}g^{LP}\partial_{L}\sigma\partial_{P}\sigma\delta_{M}^{K},
 \end{eqnarray}
 and $B_{MN}=g_{NL}B_{M}^{~L}=B_{NM}$. Also we have
 \begin{eqnarray}\label{18}
\bar{R}_{MN}=R_{MN}-g_{MN}B_{S}^{~S}-(d-2)B_{MN},
 \end{eqnarray}
 and
  \begin{eqnarray}\label{19}
\bar{g}_{MN}\bar{R}=(R-2(d-1)B_{S}^{~S})g_{MN},
 \end{eqnarray}
where $d=dim M$. Under the conformal transformation of metric, one can obtain the following results for the transformed Hermitian Ricci tensor and Ricci scalar 
 \begin{eqnarray}\label{20}
\bar{R}^{*}_{MN}=R^{*}_{MN}-B_{MN}-g_{KS}B_{R}^{K}J^{~R}_{M}J^{~S}_{N},
 \end{eqnarray}
 and
  \begin{eqnarray}\label{21}
\bar{g}_{MN}\bar{R}^{*}=g_{MN}(R^{*}-2B_{R}^{R}).
 \end{eqnarray}
\section{Appendix B}

Considering the $\Omega_{MN}=g_{NK}J_{M}^{~K}$ field as a natural generalization of the Maxwell vector field $A_{M}$, the field strength $H_{MNK}$ associated with $\Omega_{MN}$ is defined by
\begin{eqnarray}
H_{MNK}=\partial_{M}\Omega_{NK}+\partial_{N}\Omega_{KM}+\partial_{K}\Omega_{MN}=\nabla_{M}\Omega_{NK}+\nabla_{N}\Omega_{KM}+\nabla_{K}\Omega_{MN},
\end{eqnarray}
with respect to Levi-civita connection. So a non-vanishing $H_{MNK}$ needs a non-K\"{a}hler geometry. Particularly, for a strictly nearly K\"{a}hler structure the $H_{MNK}$ has the form of
\begin{eqnarray}\label{90}
H_{MNK}^{(nk)}=3~\nabla_{M}\Omega_{NK}=3~g_{KR}\nabla_{M}J_{N}^{~R}.
\end{eqnarray}
Then, in nearly K\"{a}hler geometry the $H_{MNK}H^{MNK}$ giving the dynamics to the $\Omega_{MN}$ field  will be in its \textit{simplest form}
\begin{eqnarray}
H_{MNK}^{(nk)}H^{(nk)~MNK}=9(\nabla^{K}J^{~M}_{N})(\nabla_{K}J^{~N}_{M}).
\end{eqnarray}
Obviously, the right hand side is the last term in the action \eqref{24}. The action \eqref{24} is analogues to the string effective action (within Einstein
frame) in nearly K\"{a}hler case in which the dynamics 
of the Kalb-Romond field (here $ \Omega_{MN}$) is given by a $H_{MNK}H^{MNK}$ term \cite{string}. 
\section{Appendix C}
In this section, we are going to systematically identify nearly K\"{a}hler complex structure with associated Hermitian metric.
We will write the formalism in the non-coordinate basis which is related to coordinate basis by vielbein. The covariant derivative of vielbein is given by
  \begin{eqnarray}\label{49}
  \nabla_{M}e^{N}_{~~A}=e_{M}^{~~D}e^{N}_{~~C}~\Gamma_{DA}^{C}.
  \end{eqnarray}
Using the basic definition of Levi-Civita connection and Riemannian curvature tensor, we have explicit expressions of them in non-coordinate basis in terms of structure constants and time derivation of metric as follows
 \begin{eqnarray}\label{37}
\Gamma_{AB}^{C}=\frac{1}{2}(g^{DC}(e_{A}(g_{BC})+e_{B}(g_{CA})-e_{C}(g_{AB}))-g^{DC}(f_{AC}^{~~~E}g_{EB}+f_{BC}^{~~~E}g_{EA})+f_{AB}^{~~~C}),
 \end{eqnarray}
\begin{eqnarray}\label{38}
R_{AB}=e_{D}(\Gamma_{BA}^{D})-e_{B}(\Gamma_{DA}^{D})+\Gamma_{BA}^{E}\Gamma_{DE}^{D}-\Gamma_{DA}^{E}\Gamma_{BE}^{D}-f_{DB}^{~~~E}\Gamma_{EA}^{D}.
\end{eqnarray}
Now, by employing \eqref{49} and \eqref{37}, one can obtain the nearly K\"{a}hler condition $(5a)$ in non-coordinate basis in the following form
\begin{eqnarray}\label{39}
\Gamma_{EA}^{C}J_{D}^{~A}-\Gamma_{ED}^{A}J_{A}^{~C}+\Gamma_{DA}^{C}J_{E}^{~A}-\Gamma_{DE}^{A}J_{A}^{~C}+e_{D}(J_{E}^{~C})+e_{E}(J_{D}^{~C})=0,
\end{eqnarray}
where $e_{D}$ acts on $J_{B}^{~A}$ and $g_{AB}$ as $\delta^{0}_{D}\frac{d}{dt}$. In this basis, 4-dimensional part of almost complex structure, $J_{d}^{~a}$, depends on time variable but the 6-dimensional part of it, has constant components only. Noting the fact that 6-dimensional Lie group  $SU(2)\times SU(2)$ is nearly K\"{a}hler while 4-dimensional $R\times B$ Lie group is K\"{a}hlerian\footnote{Note
that $B$ is a three dimensional Bianchi type Lie group.}, the above equation results in two decomposed equations as follows
\begin{eqnarray}\label{40}
\Gamma_{\hat{m}\hat{a}}^{\hat{n}}J_{\hat{d}}^{~\hat{a}}-\Gamma_{\hat{m}\hat{d}}^{\hat{a}}J_{\hat{a}}^{~\hat{n}}+\Gamma_{\hat{d}\hat{a}}^{\hat{n}}J_{\hat{m}}^{~\hat{a}}-\Gamma_{\hat{d}\hat{m}}^{\hat{a}}J_{\hat{a}}^{~\hat{n}}=0,
\end{eqnarray}
\begin{eqnarray}\label{41}
\Gamma_{ma}^{n}J_{d}^{~a}-\Gamma_{md}^{a}J_{a}^{~n}+e_{d}(J_{m}^{~n})=0.
\end{eqnarray}
It is useful to introduce two kinds of matrix for instance for an appeared $\Gamma_{\hat{a}\hat{b}}^{\hat{c}}$ in equations as follows
\begin{eqnarray}\label{42}
(\Gamma 1_{\hat{a}})_{\hat{b}}^{~~\hat{c}}=\frac{1}{2}(-\chi_{\hat{a}}+g.\chi_{\hat{a}}^{t}.g^{-1}+{\cal {Y}}^{\hat{d}}.g^{-1} g_{\hat{d}\hat{a}}),\nonumber\\
(\Gamma 2_{\hat{b}})_{\hat{a}}^{~~\hat{c}}=\frac{1}{2}(\chi_{\hat{b}}+g.\chi_{\hat{b}}^{t}.g^{-1}+{\cal {Y}}^{\hat{d}}.g^{-1} g_{\hat{d}\hat{b}}),
\end{eqnarray}
where $(\chi_{\hat{a}})_{\hat{b}}^{~~\hat{c}}=-f_{\hat{a}\hat{b}}^{~~\hat{c}}$ and $({\cal {Y}}^{\hat{c}})_{\hat{a}\hat{b}}=-f_{\hat{a}\hat{b}}^{~~\hat{c}}$ are adjoint representation of the Lie algebra of $su(2)\bigoplus su(2)$. Therefore, the matrix form of nearly K\"{a}hler equation \eqref{40} will be
\begin{eqnarray}\label{43}
J.\Gamma 1_{\hat{a}}-\Gamma 1_{\hat{a}}.J+\Gamma 2_{\hat{d}}* J_{\hat {a}}^{~~\hat{d}}-\Gamma 2_{\hat{a}}.J=0.
\end{eqnarray}
Now, solving the equation  \eqref{43} by using \eqref{1} and \eqref{2} for $SU(2)\times SU(2)$ gives an example of 6-dimensional nearly K\"{a}hler structure up to an arbitrary constant $\xi$ in the following form
\begin{eqnarray}\label{44}
g_{\hat{a}\hat{b}}=- 2\,\xi \left[ \begin {array}{cccccc} 1&0&0&1/2&0&0\\ \noalign{\medskip}0&1&0
&0&1/2&0\\ \noalign{\medskip}0&0&1&0&0&1/2\\ \noalign{\medskip}1/2&0&0
&1&0&0\\ \noalign{\medskip}0&1/2&0&0&1&0\\ \noalign{\medskip}0&0&1/2&0
&0&1\end {array} \right]_,
\end{eqnarray}

\begin{eqnarray}\label{444}
J_{\hat{a}}^{~\hat{b}}=\frac{\sqrt{3}}{3} \left[ \begin {array}{cccccc} -1&0&0&-2&0&0\\ \noalign{\medskip}0&-1&0
&0&-2&0\\ \noalign{\medskip}0&0&-1&0&0&-2\\ \noalign{\medskip}2&0&0&1&0
&0\\ \noalign{\medskip}0&2&0&0&1&0\\ \noalign{\medskip}0&0&2&0&0&1
\end {array} \right]_.
\end{eqnarray}
The resulted $J_{\hat{a}}^{~\hat{b}}$ and $g_{\hat{a}\hat{b}}$ with our method are in accordance with half flat structure on $S^{3}\times S^{3}$ which is obtained in  \cite{halfflat}, noting the fact that nearly K\"{a}hler structure is a special class of half flat structures.
The vielbein on $S^{3}\times S^{3}$ part is given by
\begin{eqnarray}\label{vs}
e_{\hat{\mu}}^{~~\hat{a}}= \left[ \begin {array}{cccccc} \cos \left( x_{{6}} \right) \cos
 \left( x_{{7}} \right) &-\cos \left( x_{{6}} \right) \sin \left( x_{{
7}} \right) &\sin \left( x_{{6}} \right) &0&0&0\\ \noalign{\medskip}
\sin \left( x_{{6}} \right) &\cos \left( x_{{7}} \right) &0&0&0&0
\\ \noalign{\medskip}0&0&1&0&0&0\\ \noalign{\medskip}0&0&0&\cos
 \left( x_{{9}} \right) \cos \left( x_{{10}} \right) &-\cos \left( x_{
{8}} \right) \sin \left( x_{{10}} \right) &\sin \left( x_{{9}}
 \right) \\ \noalign{\medskip}0&0&0&\sin \left( x_{{9}} \right) &\cos
 \left( x_{{10}} \right) &0\\ \noalign{\medskip}0&0&0&0&0&1
\end {array} \right]_. 
\end{eqnarray}
On the other hand, we are looking for a K\"{a}hler structure on 4-dimensional manifold. As a special example,
we consider Bianchi type \textit{II} as 3-dimensional Lie group $B$ with non-zero commutation relation \cite{Landau}
\begin{eqnarray}
[X_{2},X_{3}]=X_{1},
\end{eqnarray}
and vielbein $e_{\alpha}^{~~i}(x)$ as  \cite{mr}
\begin{eqnarray}\label{vII}
e_{\alpha}^{~~i}(x)=\left[ \begin {array}{ccc} 1&0&0\\ \noalign{\medskip}{\it x_{3}}&1&0
\\ \noalign{\medskip}0&0&1\end {array} \right]_.
\end{eqnarray}
Now, solving \eqref{41} by using \eqref{1} and \eqref{2} on $R\times II$ gives K\"{a}hler almost complex structure and Hermitian metric as follows
  \begin{eqnarray}\label{55}
 g_{ab}= \left[ \begin {array}{cccc} -{\frac {d}{dt}}F \left( t \right) &0&0&0
 \\ \noalign{\medskip}0&-{{\it c_{3}}}^{2}{\frac {d}{dt}}F \left( t
  \right) &0&0\\ \noalign{\medskip}0&0&{\it c_{1}}+F \left( t \right) {
 \it c_{2}}&0\\ \noalign{\medskip}0&0&0&{\frac { \left( {\it c_{1}}+F \left(
 t \right) {\it c_{2}} \right) {{\it c3}}^{2}}{{{\it c_{2}}}^{2}}}
 \end {array} \right]_,
  \end{eqnarray}
\begin{eqnarray}\label{555}
J_{a}^{~b}=\left[ \begin {array}{cccc} 0&-{\frac {{\it c_{3}}\,{\frac {d}{dt}}F
 \left( t \right) }{{\it c_{1}}+F \left( t \right) {\it c_{2}}}}&0&0
\\ \noalign{\medskip}-{{\it c_{3}}}^{-1}&0&0&0\\ \noalign{\medskip}0&0&0&
-{\frac {{\it c_{3}}\, \left( {\it c_{1}}+F \left( t \right) {\it c_{2}}
 \right) }{ \left( {\frac {d}{dt}}F \left( t \right)  \right) {\it c_{2}}
}}\\ \noalign{\medskip}0&0&-{\frac {{\it c_{2}}}{{\it c_{3}}}}&0\end {array}
 \right]_,
\end{eqnarray}
where $F(t)$ is an arbitrary well defined function of $t$ and $c_{1}, c_{2}$ and $c_{3}$ are arbitrary constants.


\end{document}